\documentclass[showpacs]{revtex4}
\usepackage{amsmath,amssymb}
\usepackage{natbib}
\usepackage{url}
\usepackage{graphicx}
\usepackage{psfrag}
\usepackage{color}

\usepackage{mathrsfs}
\DeclareSymbolFontAlphabet{\mathrsfs}{rsfs}
\DeclareMathAlphabet{\mathcal}{OMS}{cmsy}{m}{n}

\newcommand{\scri}{\mathrsfs{I}}
\newcommand{\be}{\begin{equation}}
\newcommand{\ee}{\end{equation}}

\def\balpha{\bar{\alpha}}
\def\bbeta{\bar{\beta}}
\def\bJ{\bar{J}}
\def\bpsi{\bar{\psi}}
\def\bpi{\bar{\pi}}

\begin{document}
\preprint{AEI-2008-081}
\title{Gravitational perturbations of Schwarzschild spacetime at null
  infinity and the hyperboloidal initial value problem}

\author{An{\i}l Zengino\u{g}lu$^{1,2,3}$, Dar{\'i}o N\'u\~nez$^{3,4}$,
         Sascha Husa$^{3,5}$} 

\affiliation{$^{1}$ Department of Physics, University of Maryland,
  College Park, MD 20742, USA}

\affiliation{$^{2}$ Center for Scientific Computation and Mathematical
  Modeling, University of Maryland, College Park, MD 20742, USA}

\affiliation{$^{3}$ Max-Planck-Institut f\"ur Gravitationsphysik,
  Albert-Einstein-Institut, 14476 Golm, Germany}

\affiliation{$^{4}$ Instituto de Ciencias Nucleares, Universidad
  Nacional Aut\'onoma de M\'exico, A.P. 70-543, M\'exico D.F. 04510,
  M\'exico,}

\affiliation{$^{5}$ Departament de F\'{\i}sica, Universitat de les Illes
Balears, Cra.~Valldemossa Km.~7.5, E-07122 Palma de Mallorca, Spain}

\begin{abstract}
We study gravitational perturbations of Schwarzschild spacetime by
solving a hyperboloidal initial value problem for the Bardeen-Press
equation. Compactification along hyperboloidal surfaces in a
scri-fixing gauge allows us to have access to the gravitational
waveform at null infinity in a general setup. We argue that this
hyperboloidal approach leads to a more accurate and efficient
calculation of the radiation signal than the common approach where a
timelike outer boundary is introduced. The method can be
generalized to study perturbations of Kerr spacetime using the
Teukolsky equation.
\end{abstract}

\maketitle

\section{Introduction}

In the last few years, numerical relativity has made tremendous
progress in treating the inspiral of compact objects, starting with
the breakthrough simulations of
\cite{Pretorius:2005gq,Campanelli:2005dd,Baker:2005vv}. Rather
accurate simulations of the last $\approx 10$ orbits of roughly equal
mass binary black holes have become standard in the field (see
e.g. \cite{Hannam:2007wf, Hinder:2008kv, Campanelli:2008nk,
  Boyle:2008ge, Baker:2008mj,Pollney:2007ss} for recent
examples). However, these simulations are still restricted to finite
outer boundaries and gravitational radiation is extracted at a number
of finite shells. Maybe not surprisingly, it has been found that the
wave extraction radius is a significant limiting factor for the
precision of results \cite{Pazos2006, Hannam:2007ik,Boyle:2007ft,
  Kocsis:2007zz}.  The practical implementation of algorithms that
include null infinity in the numerical grid is thus highly
desirable. It also provides for a fruitful problem to foster the
interaction of numerical and mathematical relativity.

An important stepping stone to develop such algorithms is the case of
linearized perturbations on a fixed background. The numerical study of
linearized perturbations is convenient as one can evolve the
gravitational radiation very accurately for a comparatively low
computational cost
\cite{Campanelli:1998uh,Baker:2001sf,PazosAvalos:2004rp}.  For
evolutions along null surfaces this has provided important guidance in
the past, see
e.g. \cite{Husa:2001pk,Campanelli:2000in,Zlochower:2003yh}.  Note,
however, that linear analysis may not be valid in certain special weak
field situations. Examples where the linear analysis breaks down have
been found in studies of test fields
\cite{Bizon:2007qz,Bizon:2007xa}. In the pure gravitational case, so
far it seems that the perturbative analysis fits well with the
nonlinear evolution \cite{Baker:1999sj, Baker:2001sf, Baker:2002qf,
  Baker:2003ds, Campanelli:2005ia}.

When the background spacetime describes the particularly interesting
case of a stationary black hole, the perturbation equations take the
form of linear wave equations with a potential term (in our case the
Bardeen-Press equation \cite{Bardeen:1973xb}).  The solutions to such
equations show three stages: an initial transient (the ``direct
signal''), quasi-normal mode ringing and power-law decay. The standard
method to numerically analyze these stages in the time domain has been
to solve initial boundary value problems using Cauchy-type
foliations. This method has certain disadvantages. When one is
primarily interested in radiative properties of the solution far away
from the source, the use of Cauchy-type foliations is a waste of
computational resources. Furthermore, there are difficulties related
to the choice of boundary conditions. It has been shown that a typical
choice of boundary data for numerical relativity destroys the
polynomial tail behavior of solutions to wave equations on a
Schwarzschild spacetime \cite{Allen:2004js,Dafermos:2004wt}. Note that
even if the polynomial tail can be calculated by using a better choice
of boundary data, the decay rate will be the one near time-like
infinity. It has been pointed out, however, that the relevant decay
rate for an astrophysical observer is the one along null infinity
\cite{Leaver:1986gd,Purrer:2004nq,Zeng08a}.

Using foliations that approach null infinity would cure these
problems. Numerical studies of gravitational perturbations at null
infinity have been performed by \cite{Husa:2001pk,Campanelli:2000in}
using a characteristic approach for the Bardeen-Press equation, and
have been compared with fully nonlinear studies in
\cite{Zlochower:2003yh}. The authors have been able to efficiently
calculate the quasi-normal mode ringing and the power law decay. There
are, however, two main difficulties in the characteristic approach. An
important aspect of solving the Bardeen-Press equation instead of the
Regge-Wheeler-Zerilli equations is that the favorable separability
properties of the Teukolsky formalism extends to Kerr spacetime
\cite{Teu73}.  Unfortunately, generalizing the studies of
\cite{Husa:2001pk,Campanelli:2000in} to Kerr spacetime is complicated
because the construction of null foliations smoothly covering Kerr
spacetime is not straightforward and from a numerical point of view
not very convenient \cite{Bishop05,Pretorius98}. A further
complication is related to the causal structure of a black hole
spacetime. Due to the global relationship between null surfaces on
which boundary information is known to pose a characteristic initial
value problem, the numerical solution for a perturbed black hole needs
to be calculated in two steps: advanced \cite{Campanelli:2000in} and
retarded \cite{Husa:2001pk}. A related difficulty shows itself also in
the choice of suitable coordinates which allow the computation of the
gravitational waveform at null infinity and provide good resolution
near the horizon.

An alternative, and very flexible method to study radiation is to
use spacelike surfaces that approach null infinity. Following
\cite{Friedrich83a} we call such surfaces hyperboloidal and the
related initial value problem a hyperboloidal initial value
problem. Hyperboloidal surfaces are as flexible as Cauchy-type
surfaces because the property that characterizes them is not local but
global. Further, they enable, like characteristic surfaces, a clean
treatment of gravitational radiation because they approach null
infinity. In this sense, they are hoped to combine the ``best of both
worlds'', and we expect increasing interest in this approach.

In the present paper we want to learn from the linearized case as a
preparation for the nonlinear hyperboloidal initial value problem. To
this end we construct solutions of the Bardeen-Press equation by
numerically solving a hyperboloidal initial value problem for the
perturbations of Schwarzschild spacetime and we calculate the
gravitational wave signal at null infinity. We find that
compactification along hyperboloidal surfaces leads to a very
efficient code. Specifically, we show that the calculation of the
quasi-normal mode frequencies in the time domain can be performed
accurately with a small number of grid points using standard numerical
techniques. The last stage of the solution, namely the power-law
decay, can be numerically calculated if the resolution is sufficiently
high. A major advantage of our method for this part of the signal is
not so much the efficiency but rather the fact that we can measure the
decay rate at null infinity by solving a Cauchy problem in the PDE
sense.

In the presentation and, to some extent, in the choice of the methods
applied in the paper, we tried to avoid a heavy use of the conformal
language. Our philosophy of constructing an algorithm for the
hyperboloidal initial value problem follows the suggestions in
\cite{Andersson:2002gn, Husa:2005ns}. Instead of constructing a
gauge-independent, regular, conformal Bardeen-Press equation, we make
a suitable use of our gauge freedom before compactifying the
coordinates and show that, in the end, regular equations are
obtained. While both approaches are equivalent in their final aim, we
hope that the way we present the method will be more accessible to
numerical relativists who are not familiar with notions related to
conformal geometry.

The paper is organized as follows: In section II we present the
Teukolsky formalism and derive the Bardeen-Press equation for a
spherically symmetric but otherwise arbitrary foliation. In the third
section we discuss our method to include null infinity on a numerical
grid. We show that the hyperboloidal Bardeen-Press equation written in
compactifying coordinates is regular at null infinity. After a short
presentation of our numerical methods in section IV, we go over to the
numerical studies of the last two stages for the gravitational
perturbations where we demonstrate some of the basic advantages of the
hyperboloidal approach in numerical work. In the discussion section we
summarize our results and point out some ideas for future work.

\section{The perturbation equation}

\subsubsection{General definitions}

Teukolsky \cite{Teu73} derived a master perturbation equation for
various fields in a Kerr background, using the spinor formalism
introduced by Newman and Penrose \cite{NP62}. One of the central ideas
of this formulation consists in choosing a tetrad of null vectors,
${Z_a}^\mu$, and then defining the directional operators as the
projections of the covariant derivative along each null vector.
For a complete review on the subject and its properties, the reader is
referred to \cite{Chandra83, Frolov77}.

Usually the tetrad is defined with four null vectors: two real and
along the light cone, the other two complex conjugate and in the
perpendicular plane to the cone. There is much room for conventions
regarding the orientation and scaling of the null vectors, and
consequently for confusion. We make the following choices: we label the
two real vectors along the light cone as ${Z_0}^\mu=l^\mu,
{Z_1}^\mu=k^\mu$ and choose them both future directed with $l^\mu$
pointing outward and $k^\mu$ pointing inward.
The complex two null vectors will be denoted by $m^\mu$:
${Z_2}^\mu=m^\mu, {Z_3}^\mu={m^*}^\mu$.

After these choices, there is still the issue of the normalization of
the products $l^\mu\,k_\mu$, and $m^\mu\,{m^*}_\mu$. The rest of the
products are zero by construction. The properties mentioned above can
be expressed in the following equations:
\begin{equation}
{Z_a}^\mu\,{Z_b}_\mu=\eta_{ab}, \hspace{1cm} g_{\mu\nu}=2\eta^{ab}\,Z_{a\,(\mu}\,Z_{|b|\,\nu)},\label{eqs:tetrad}
\end{equation}
with $\eta_{ab}$ a matrix of the form
\begin{equation}
\eta_{ab}=\left(\begin{matrix}&0&\,
  &\eta_{12}&\,&0&\,&0&\\&\eta_{12}&\,&0&\,&0&\,&0&\\&0&\,&0&\,&0&\,&-\eta_{12}&\\&0&\,&0&\,&-\eta_{12}&\,&0&\end{matrix}\right),
\end{equation}
with $\eta_{12}=\pm 1$ depending on the signature.  If we choose the
signature of the spacetime to be $(+,-,-,-)$ as in \cite{Chandra83,
  Frolov77}, then the choice consistent with Eqs.~(\ref{eqs:tetrad})
to derive the perturbation equation is $\eta_{12}=1$ \cite{Teu73}.  If
the spacetime is described using the signature $(-,+,+,+)$ we have to
choose $\eta_{12}=-1$ in order to remain consistent with
Eqs.~(\ref{eqs:tetrad}). Some discussion on the change of the tetrad
due to signature can be found in \cite{Ernst78}.

We derived the linearized equation for the perturbed Weyl scalar
$\Psi_{4}$ (note that the background value vanishes), leaving
unspecified the choice of signature. Using the definitions for the
directional operators and spinor coefficients as given in
\cite{Chandra83, Frolov77}, the final perturbation equation takes the
form
\begin{eqnarray}
&&\left(\left({\bf \Delta} + \eta_{12}\, \left(4\,\mu_s + {\mu_s}^* + 3\,
  \gamma_s -{\gamma_s}^* \right)\right)\,\left({\bf D} -
  \eta_{12}\,\left(\rho_s -4\,\epsilon_s \right) \right) - \left({\bf
    \delta^*} + \eta_{12}\, \left(3\,\alpha_s + {\beta_s}^* + 4\,\pi_s - {\tau_s}^*
  \right)\right)\,\left( {\bf \delta} + \eta_{12}\,\left(4\,\beta_s-\tau_s
  \right) \right) \right.  \nonumber \\ -
  &&\left. 3\,\eta_{12}\,\Psi_2 \right)\,\Psi_{4}
  =0 \label{eq:PertPsi4f}
\end{eqnarray}
We have added a subindex ${}_s$ to the spinor coefficients in order to avoid
confusion with other quantities that will be used later. In the rest of the work
we will be using the $(-,+,+,+)$ signature, and thus $\eta_{12}=-1$.

\subsection{Spherical symmetry}

We will work in spherically symmetric static spacetimes. Then the
metric can be written as
\be \label{eq:decomp} g = (-\alpha^2+\gamma^2\beta^2)\,dt^2 + 2
\gamma^2 \beta\, dt\,dr + \gamma^2\,dr^2+r^2\,d\sigma^2.  \ee We will
write the Bardeen-Press equation for $\Psi_4$ on such a background
with respect to an orthonormal null tetrad $(l,k,m,\bar{m})$,
consistent with the conditions mentioned above.  The tetrad can be
given by
\[l^\mu=\frac{1}{2\alpha^2}\left(1,\frac{\alpha}{\gamma}-\beta,0,0\right),
\qquad
k^\mu=\left(1,-\left(\frac{\alpha}{\gamma}+\beta\right),0,0\right),
\qquad m^\mu=\frac{1}{\sqrt{2}r}(0,0,1,i \csc \theta), \qquad
\bar{m}^\mu=(m^\mu)^\ast.\]
Remember that $\Psi_4$ is defined by contracting the Weyl tensor with
respect to the inward pointing null vector field of the null tetrad,
$k$ in our case.

We compute the corresponding spinor coefficients and directional
derivatives and substitute into the perturbation equation
(\ref{eq:PertPsi4f}). Using the definitions for the spinor
coefficients given in \cite{Chandra83}, and denoting derivatives with
respect to $r$ by a prime, we obtain the non zero spinors and Weyl
scalars as
\begin{eqnarray}
&\rho_s=\frac{\frac{\alpha}{\gamma}-\beta}{2\,r\,\alpha^2}; \hspace{0.5cm}
  \mu_s=\frac{\frac{\alpha}{\gamma}+\beta}{r}; \hspace{0.5cm}
  \epsilon_s=-\frac{\left(r\,\alpha\,\gamma\,\rho_s\right)^\prime}{2\,\alpha\,\gamma}\,; \hspace{0.5cm}
  \gamma_s=-\frac{\left(r\,\alpha\,\gamma\,\mu_s\right)^\prime}{2\,\alpha\,\gamma};&
  \nonumber
  \\ &\beta_s=-\frac{\cot\theta}{2\,\sqrt{2}\,r}; \hspace{0.5cm}
  \alpha_s=-\beta_s, \hspace{0.5cm}
  \Psi_2=\frac{\left(r^2\,\alpha\,\gamma\,\rho_s\,\mu_s\right)^\prime}{r\,\alpha\,\gamma}.& \label{coef_spinor}
\end{eqnarray}

Before proceeding to write down the perturbation equation for
$\Psi_4$, we first note that the simple sourceless wave equation,
\begin{equation}
g^{\mu\nu}\,\nabla_\mu\,\nabla_\nu\,\Phi(x^\alpha)=0,
\end{equation}
independently of the tensorial or spinorial character of $\Phi$
in this background takes the form
\[ \left[{{\cal O} }_{rt} + {{\cal O} }_{\theta\varphi}\right]\,\Phi=0, \]
with
\begin{eqnarray}
{{\cal O} }_{rt}&=&\frac{r^2}{\alpha^2}\left(-\frac{\partial^2}{\partial t^2} + 2\,\beta\,\frac{\partial^2}{\partial t\,\partial r} + \left(\left(\frac{\alpha}{\gamma}\right)^2-\beta^2\right)\frac{\partial^2}{\partial r^2} + \frac{\alpha}{r^2\,\gamma}\left(\frac{r^2\,\beta\,\gamma}{\alpha}\right)^\prime\frac{\partial}{\partial t} + \frac{\alpha}{r^2\,\gamma}\left(r^2\,\left(\frac{\alpha}{\gamma} -\frac{\beta^2\,\gamma}{\alpha} \right)\right)^\prime\frac{\partial}{\partial r}\right), \\
{{\cal O} }_{\theta\varphi}&=& \frac{\partial^2}{\partial \theta^2} + \frac1{\sin^2\theta}\,\frac{\partial^2}{\partial \varphi^2} + \cot\theta\,\frac{\partial}{\partial \theta}.
\end{eqnarray}

In the perturbation equation (\ref{eq:PertPsi4f}) we use the sign
consistent with our choice of signature, that is $\eta_{12}=-1$.
After substituting our tetrad and the obtained directional operators,
spinor coefficients and the Weyl scalar $\Psi_2$, we obtain
\begin{equation}
\left[L_{tr} + L_{\theta\varphi} \right]\,\Psi_{4}=0\,, \label{eq:pertPsi4KS}
\end{equation}
where the operators $L_{tr}$ and $L_{\theta\varphi}$ are given by
\begin{eqnarray}
\label{eq:radBP} L_{tr}&=&-\frac{r^2}{\alpha^2}\frac{\partial^2}{\partial t^2} + 2\,\beta\,\frac{r^2}{\alpha^2}\,\frac{\partial^2}{\partial t\,\partial r} + r^2\,\left(\frac{\alpha^2-\beta^2\,\gamma^2}{\alpha^2\,\gamma^2}\right)\frac{\partial^2}{\partial r^2} +  A\,\frac{\partial}{\partial t} + B\,\frac{\partial}{\partial r} - 2\,r^2\,C, \\
\label{eq:angular}L_{\theta\varphi}&=&{{\cal O} }_{\theta\varphi}  -4\,i\,\frac{\cos\theta}{\sin^2\theta}\frac{\partial}{\partial \varphi} -2\,\frac{\cos^2\theta + 1}{\sin^2\theta}.
\end{eqnarray}
Here we have defined
\begin{eqnarray}
A&=&\frac1{\alpha\,\gamma}\left(\frac{r^2\,\beta\,\gamma}{\alpha}\right)^\prime-4\frac{r^4\,\mu_s}{\alpha^3}\,\left(\frac{\alpha}{r}\right)^\prime, \nonumber \\
B&=&\frac2{\alpha\,\gamma}\left(r^4\,\alpha\,\gamma\,\rho_s\,\mu_s\right)^\prime - 4\,r^2\,\left(\frac{1}{\gamma^2}\,\left(\frac{\beta\,\gamma}{\alpha} \right)^\prime + 2r^3\,\frac{\rho_s\,\mu_s}{\alpha}
\left(\frac{\alpha}{r}\right)^\prime \right), \nonumber\\
C&=&\frac{1}{r\,\alpha\,\gamma}\,\left(3\,\left(\alpha\,\gamma\,\rho_s\,\mu_s\,r^2\right)^\prime+\rho_s\,r\,\left(\alpha\,\gamma\,\mu_s\,r\right)^\prime - 10\,\mu_s\,r\,\left(\alpha\,\gamma\,\rho_s\,r\right)^\prime\right) -\mu_s\,\left(5\,\rho_s + r\,{\rho_s}^\prime + 2\,r\,\left(\frac{\left(\alpha\,\gamma\,\rho_s\,r\right)^\prime}{\alpha\,\gamma}\right)^\prime \right) \nonumber \\ && + \frac{2}{\alpha^2\,\gamma^2}\,\left(\alpha\,\gamma\,\rho_s\,r\right)^\prime\,\left(\alpha\,\gamma\,\mu_s\,r\right)^\prime,
\label{coef_fo}
\end{eqnarray}
There are several things worth noticing in this equation. Its
principal part is the same as of the wave equation. The gravitational
contribution is described by some extra coefficients in the terms
containing no higher than first radial derivatives. The left hand side
of the equation for the gravitational perturbation, as well as that
for the wave equation, is separable into terms involving the $(r, t)$
and the angular coordinates. The equation is somewhat lengthy, due to
the fact that we are considering general metric coefficients, but
nevertheless, it is quite tractable. Finally, we remark the fact that
had we not considered the corresponding change in signature, we would
have obtained an inconsistent expression.

We will work with the function $\widetilde{\Psi}_4=r\,\Psi_{4}$, in order to
factor out the expected decay due to the peeling behavior. The
function $\widetilde{\Psi}_4$ satisfies the equation:
\begin{equation}
\left[\widetilde{L}_{tr} + L_{\theta\varphi}
  \right]\,\widetilde{\Psi}_4=0, \label{eq:pertPhi1KS}
\end{equation}
where the operator $\widetilde{L}_{tr}$ reads
\begin{equation}
\widetilde{L}_{tr}=-\frac{r^2}{\alpha^2}\,\frac{\partial^2}{\partial
  t^2} + 2\,\beta\,\frac{r^2}{\alpha^2}\,\frac{\partial^2}{\partial
  t\,\partial r} + 2\,r^4\,\rho_s\,\mu_s\,\frac{\partial^2}{\partial
  r^2} + \widetilde{A}\,\frac{\partial}{\partial t} +
\widetilde{B}\,\frac{\partial}{\partial r} -
2\,r^2\,\widetilde{C}, \label{eq:optrPhi1}
\end{equation}
where now
\be
\widetilde{A}=A -2\,r\,\frac{\beta}{\alpha^2}, \qquad
\widetilde{B}=B - 4\,r^3\,\rho_s\,\mu_s, \qquad 
\widetilde{C}=C + \frac{B}{2\,r^3} - 2\,\rho_s\,\mu_s, \label{coef_fo1}
\ee
and $A, B$, and $C$ are given by (\ref{coef_fo}).  Further, using the
fact that the Weyl scalar has spin weight $-2$ \cite{Chandra83} to
expand $\widetilde{\Psi}_4$ in terms of spin-weighted spherical
harmonics
\begin{equation}
\widetilde{\Psi}_4=\sum\limits_{lm}\,\phi_{l,m}(t,r)\,{Y_{-2}}^{l,m}(\theta,\varphi).
\end{equation}
To deal with spin-weighted harmonics, it is useful to introduce the so
called  ``eth'' and ``eth-bar'' operators \cite{Newm66,Gold67}:
\begin{eqnarray}
\eth_s=-\left(\frac{\partial}{\partial \theta} + i\,\csc\theta\,\frac{\partial}{\partial \varphi} - s\,\cot\theta\right)\equiv \eth_0 + s\,\cot\theta, \label{op:eth} \\
{\bar\eth}_s=-\left(\frac{\partial}{\partial \theta} - i\,\csc\theta\,\frac{\partial}{\partial \varphi} + s\,\cot\theta\right)\equiv {\bar\eth}_0 - s\,\cot\theta, \label{op:beth}
\end{eqnarray}
which have the property of raising and lowering the spin weight of the
function on which they act. It is straightforward to show that the
angular operator in 
(\ref{eq:angular}) can be expressed in terms of the eth operators as
\begin{equation}
L_{\theta\varphi}={\bar\eth}_{-1}\,\eth_{-2}.
\end{equation}
The ${Y_{-2}}^{l,m}$ are the eigenfunctions of this angular operator:
\begin{equation}
L_{\theta\varphi}\,{Y_{-2}}^{l,m}={\bar\eth}_{-1}\,\eth_{-2}\,{Y_{-2}}^{l,m}= -\left(l-1\right)\,\left(l+2\right)\,{Y_{-2}}^{l,m}. \label{eq:Ym2}
\end{equation}

In this way, the perturbation equation (\ref{eq:pertPhi1KS})
transforms into a 1+1 equation for each $(l,m)$-mode of
the gravitational perturbation,
\begin{equation}
\left[\widetilde{L}_{tr} - \left(l-1\right)\,\left(l+2\right)
  \right]\,\phi=0, \label{eq:pertPhi1KSf}
\end{equation}
where we have dropped the indices $l,m$. In our numerical work, we
have used a first order reduction of this equation as derived in the
Appendix.
\section{Including null infinity in the numerical domain}
In the previous section we followed the standard procedure to
calculate gravitational perturbations on a spherically symmetric
background spacetime using the Teukolsky formalism with the additional
feature that we allowed an arbitrary spherically symmetric foliation
of the Schwarzschild background (see \cite{Sarbach:2001qq} for the
corresponding generalization in the Regge-Wheeler-Zerilli
formalism). If we now insert the standard Schwarzschild gauge
functions into equation (\ref{eq:pertPhi1KSf}) we get the familiar
Bardeen-Press equation. Our aim, however, is to include null infinity
in the numerical domain. In this section we show how this can be done.

\subsection{Choice of a hyperboloidal foliation}
\begin{figure}[ht]
  \centering 
  \begin{minipage}[t]{0.47\textwidth}
    \flushright 
    \psfrag{ip}{$i^+$} \psfrag{scrp}{$\scri^+$}
    \psfrag{hor}{$\mathcal{H}^+$} \psfrag{sing}{\scriptsize{singularity}}
    \includegraphics[height=0.21\textheight]{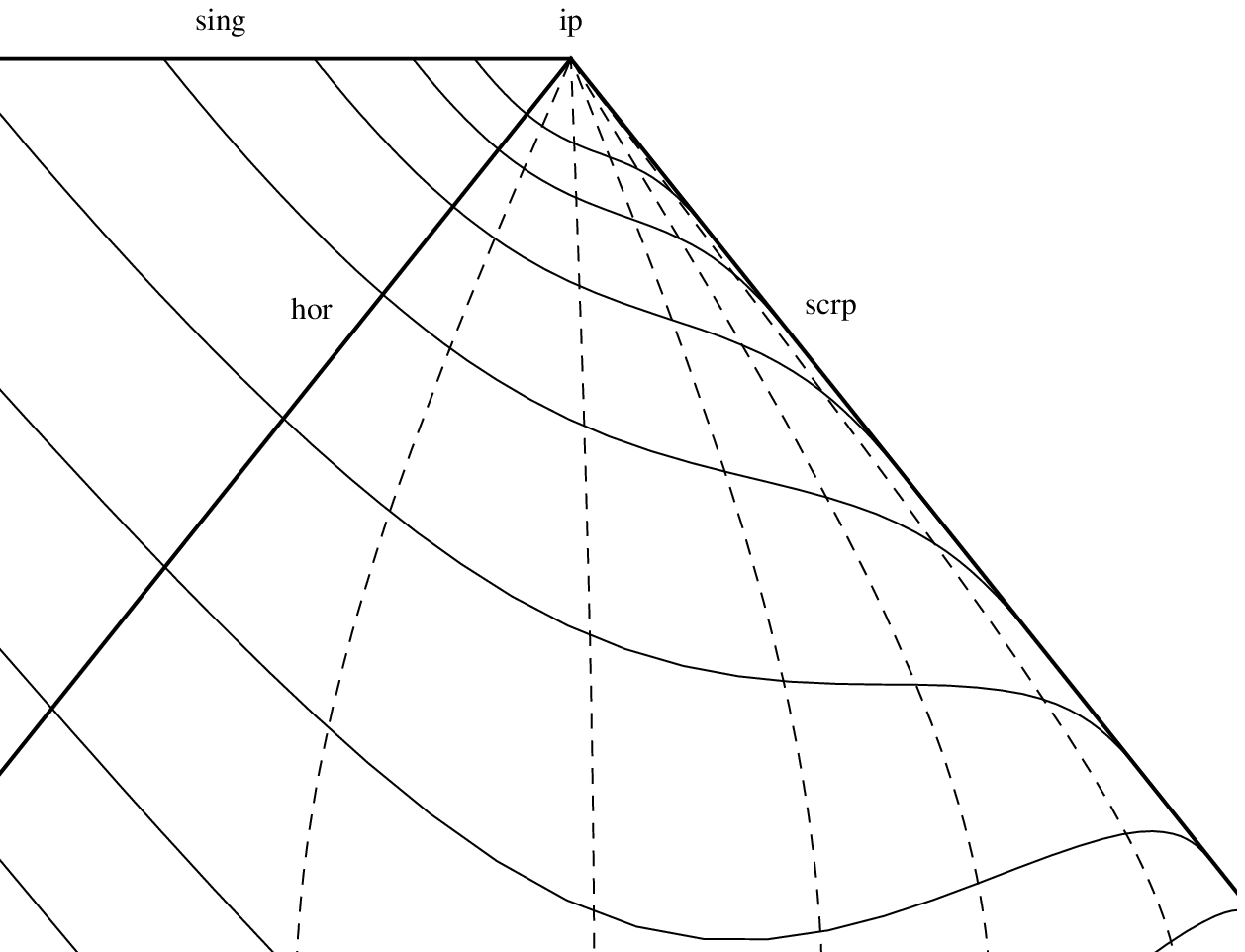}\hspace{1cm}
    \caption{Penrose diagram of a CMC-foliation in Schwarzschild
      spacetime with $m=1/2$, $c=1, K=1$. The dashed lines
      represent Killing observers. \label{fig:1}}
  \end{minipage}\hfill
  \begin{minipage}[t]{0.47\textwidth}
    \centering 
    \psfrag{c}{$\frac{A_\phi}{\Omega^2}$} \psfrag{r}{$\rho$}
    \includegraphics[height=0.2\textheight]{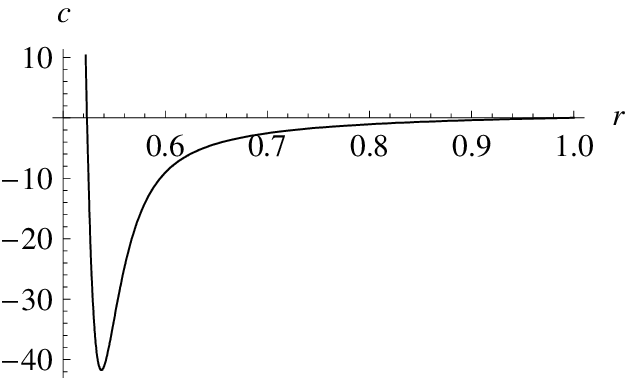}
    \caption{The function $A_\phi/\Omega^2$ from (\ref{eq:Com2}) for the
      same hyperboloidal foliation as in the diagram. We see that
      $C/\Omega^2$ vanishes at null infinity.  \label{fig:2}}
  \end{minipage}
\end{figure}

To include null infinity in the setting of a space-time splitting we
first choose a hyperboloidal foliation by introducing a new time
coordinate, $\tau$, whose level sets are hyperboloidal
surfaces. Following \cite{Zeng08a}, we consider the class of
spherically symmetric constant mean curvature (CMC) foliations of
Schwarzschild spacetime \cite{Brill80, MalecMurch03}. The
transformation from the standard Schwarzschild time coordinate to the
time coordinate of a spherically symmetric CMC-foliation can be
written as $\tau=t-h(r),$ where $r$ is the Schwarzschild area radius,
$h(r)$ is the height function and $t$ is the standard Schwarzschild
time coordinate. The height function is not known in closed form. Its
radial derivative is given by \be\label{eq:height} h'(r)=\frac{J(r)}
{\left(1-\frac{2m}{r}\right) P(r)}, \quad \mathrm{where} \quad J(r) =
\frac{K r}{3}-\frac{c}{r^2}, \quad \mathrm{and} \quad P(r) :=
\sqrt{J(r)^2+\left(1-\frac{2m}{r}\right)}.\ee Here, the mass of the
Schwarzschild black hole is denoted by $m$. The foliation parameters
are the mean extrinsic curvature $K$, and a constant of integration
$c$.  The global behavior of CMC-surfaces depends on the foliation
parameters. We choose these parameters such that the surfaces of the
foliation come from future null infinity, pass the event horizon above
the bifurcation sphere and run into the future singularity as depicted
in Figure \ref{fig:1} \cite{Zeng07}. The causal nature of the
hyperboloidal foliation shown in the conformal diagram explains why we
do not need a two stage calculation of the waveform in contrast to the
characteristic case \cite{Husa:2001pk,Campanelli:2000in}. A further
convenient feature of this foliation is that the natural coordinates
are adapted to time symmetry in the sense that the timelike Killing
vector field of Schwarzschild spacetime $\partial_t$ is given by
$\partial_\tau$ in the new coordinates. Therefore, timelike curves
with constant spatial coordinates can be regarded as worldlines of
natural observers at constant distances.

In coordinates adapted to the CMC slicing, the standard Schwarzschild
metric is obtained in the form
\[ g = -\left(1-\frac{2 m}{r}\right)\, d\tau^2 -
\frac{2J(r)}{P(r)}\,d\tau dr+ \frac{1}{P^2(r)}dr^2 +
r^2\,d\sigma^2, \] By comparing with (\ref{eq:decomp}) we see that the
gauge functions read \be\label{eq:gaugephy}
\alpha(r)=P(r),\qquad \beta(r) = -J(r)\,\alpha(r), \qquad \gamma(r) =
\frac{1}{\alpha(r)}.\ee An important difference to Cauchy foliations
should be noted at this point. The asymptotic behavior of an
asymptotically flat spacetime written in a Cauchy foliation implies
$\alpha \sim 1$ and $\beta \sim 0$ as $r\to \infty$. The same
spacetime written in a hyperboloidal foliation implies a different
asymptotic behavior for the gauge functions, namely
\be\label{eq:asympt} \alpha \sim O(r) \quad \mathrm{and} \quad \beta
\sim O(r^2) \quad \mathrm{as} \quad
r\to\infty.\ee In the next subsection, we will see that this behavior
is important for the regularity of our equations.

\subsection{Compactification}
To include the asymptotic ends of the surfaces of our hyperboloidal
foliation, we introduce a compactifying coordinate $\rho$ that is
related to the standard Schwarzschild area-radius $r$ via
\be
\label{eq:compr} r = \frac{\rho}{1-\rho} = \frac{\rho}{\Omega}
\qquad \mathrm{where} \qquad \Omega:=1-\rho.  \ee We choose the
black-hole mass as $m=1/2$, the horizon is thus located at
$\rho=1/2$. Future null infinity, $\scri^+$, is located at $\rho=1$.
Other choices are, of course, possible.  Before writing the
Bardeen-Press equation in the compactifying coordinate $\rho$, we
factor out the singular asymptotic behavior of our hyperboloidal
gauge functions (\ref{eq:asympt}) in order to deal with regular
functions. Considering the asymptotic behavior of the conformal
factor, $\Omega\sim 1/r$, we define $\balpha := \Omega \,\alpha$ and
$\bbeta := \Omega^2\beta$. The rescaled gauge source functions become
\be \label{eq:resc} \balpha =
\sqrt{\bJ^2+\left(1-\frac{2\,m\,\Omega}{\rho}\right) \Omega^2}, \qquad
\bbeta = -\bJ \,\balpha \qquad \mathrm{where} \qquad \bJ := \Omega\, J
= \frac{K\rho}{3} - \frac{c\,\Omega^3}{\rho^2}.\ee These functions are
obviously regular at $\{\Omega=0\}$. Furthermore, they satisfy
\be\label{eq:scrifix}
\left(\balpha^2-\frac{\bbeta^2}{\balpha^2}\right)= O(\Omega^2) \qquad
\mathrm{with} \qquad \beta|_{\scri^+}<0.\ee We note that this
behavior with the choice of the conformal factor (\ref{eq:compr}) is
specific to scri-fixing gauges and cannot be expected for all
hyperboloidal foliations. It is related to the fact that the normal to
level sets of $\tau$ lies in $\scri^+$ \cite{Zeng07}.

The definitions of the auxiliary variables given in (\ref{eq:aux})
imply
\[ \psi= \partial_r \phi = \Omega^2\partial_\rho\phi =: \Omega^2 \bpsi,
\quad\mathrm{and}\quad \pi =
\frac{1}{\alpha^2}(\partial_t\phi-\beta\partial_r \phi)
=\frac{\Omega^2}{\balpha^2}(\partial_t\phi-\bbeta\bpsi)
=:\Omega^2\bpi.\] We write the system (\ref{eq:firstorder}) with
respect to the functions $\balpha, \bbeta, \bpsi, \bpi$ by factoring
an $\Omega^2$ term out of the equations. The resulting first order,
symmetric hyperbolic system of equations reads
\begin{eqnarray*}
  \partial_\tau \phi &=&\balpha^2 \,\bpi + \bbeta\bpsi,
  \\ \partial_\tau \bpsi &=&\partial_\rho\left(\balpha^2 \bpi +
  \bbeta\bpsi\right), \\ \partial_\tau \bpi &=&
  \partial_\rho\left(\balpha^2\bpsi+\bbeta\bpi\right)+
  A_\pi\bpi+A_\psi\bpsi+\left(\frac{A_\phi}{\Omega^2}-
  \frac{\lambda}{\rho^2}\right)\phi.
\end{eqnarray*}
where $A_\pi, A_\psi$, and $A_\phi$ are functions of $\rho$. It turns
out that $A_\pi$ and $A_\psi$ are conformally covariant. Writing
$A_\pi$ and $A_\psi$ from (\ref{eq:ABC}) in terms of the rescaled
gauge functions and the compactifying coordinate $\rho$ results in
\[A_\pi =-4\,(\balpha^2+\bbeta)\, \partial_\rho\left(
\ln\frac{\balpha}{\rho}\right),\qquad A_\psi = A_\pi -
4\,\balpha^2\,\partial_\rho\left(\frac{\bbeta}{\balpha^2}\right).\] To
show the regularity of our equations, we need to show that $A_\phi$ falls
off with $\Omega^2$. We have by
(\ref{eq:ABC}) \begin{eqnarray} \label{eq:Com2}
  \frac{A_\phi}{\Omega^2}&=&-\frac{\Omega}{\rho}\, \partial_\rho
  \left(\frac{1}{\Omega^2} \left(\balpha^2
  -\frac{\bbeta^2}{\balpha^2}\right)\right) +
  \frac{4}{\Omega\rho}\,\left(\balpha^2\partial_\rho
  \left(\frac{\bbeta}{\balpha^2}\right) + \left(\balpha^2
  -\frac{\bbeta^2}{\balpha^2}\right)\,\partial_\rho\left(
  \ln\frac{\balpha}{\rho}\right)\right) \nonumber \\ &+&
  2\,\frac{\left(\balpha^2+\bbeta\right)^2}{\Omega^2\rho^2}\,
  \partial_\rho\left(\frac{\Omega^4\partial_\rho
    \left(\frac{\rho^2}{\Omega^2}\left(1-\frac{\bbeta}
         {\balpha^2}\right)\right)} {\balpha^2+\bbeta}\right).
\end{eqnarray} 
The explicit regularity of the above expression follows from the
behavior of our rescaled gauge functions (\ref{eq:scrifix}),
specifically from $\balpha^2+\bbeta \sim O(\Omega^2)$ near null
infinity. Indeed, as we can see in Figure \ref{fig:2} the expression
on the right hand side of (\ref{eq:Com2}) is not just regular but
vanishes at null infinity for the foliation given in
(\ref{eq:resc}). We note that scri-fixing plays an essential role
here.

\section{Numerical studies}

\subsection{Initial data}

As we are not interested in the first (transient) part of the signal,
the initial data can be chosen quite freely. We take a simple Gaussian
as the initial perturbation and set
\[ \phi(0,\rho) = a\,e^{-(\rho-\rho_c)^2/\sigma^2}, \qquad 
\psi(0,\rho)=\partial_\rho\phi(0,\rho), \qquad \pi(0,\rho) = 0. \]
Here, $\rho_c$ is the center of the Gaussian pulse, $a$ is its
amplitude and $\sigma$ is its width. We set $\rho_c=0.7, a=1$ and
$\sigma=0.05$ on the domain $\rho\in[0.495,1]$. The solid line in
Figure \ref{fig:7} shows the initial data. We have tested that
different choices of initial data do not have a relevant influence on
the features we discuss in the following. Note that the vanishing of
$\pi(0,r)$ does not mean that the data is time symmetric, because the
initial surface is future hyperboloidal (spacelike while approaching
future null infinity). We focus on the typically dominant $l=2$ mode,
corresponding to the lowest allowed angular momentum number for a
spin-2 field because higher modes are weaker due to their faster decay
rates and longer quasinormal mode ringing phases.
\subsection{Numerical algorithms}
We adopt the method of lines approach and discretize time and space
separately, using 4th order Runge-Kutta time integration and finite
differencing with 4th, 6th and 8th order accurate stencils. We add
Kreiss-Oliger type artificial dissipation to the evolution equation
for $\pi$ to suppress numerical high-frequency waves
\cite{KreissOliger73}. For a $2p-2$ accurate scheme we choose an
operator $(Q)$ of order $2p$ as
\[ Q=\ \epsilon (-1)^p \frac{h^{2p-1}}{2^p}\,D_+^p D_-^p,  \]
where $h$ is the grid size, $D_\pm$ are the forward and backward
finite differencing operators and $\epsilon$ is the dissipation
parameter. Unless otherwise stated, we set $\epsilon=0.07$.

The inner boundary is a spacelike surface inside the event horizon
where no boundary data is required. We excise the interior domain near
the singularity. For the class of CMC surfaces we have chosen,
excision is necessary as there is a minimal surface inside the
event horizon where our coordinates break down. The outer boundary is
at null infinity where one-sided finite differencing is applied. The
one-sided finite differencing is of the same order as the interior
one.

\begin{figure}[ht]
  \centering 
  \begin{minipage}[t]{0.47\textwidth}
    \center
    \psfrag{t}{$\tau/m$}  \psfrag{Q}{$Q$}
    \includegraphics[height=0.17\textheight]{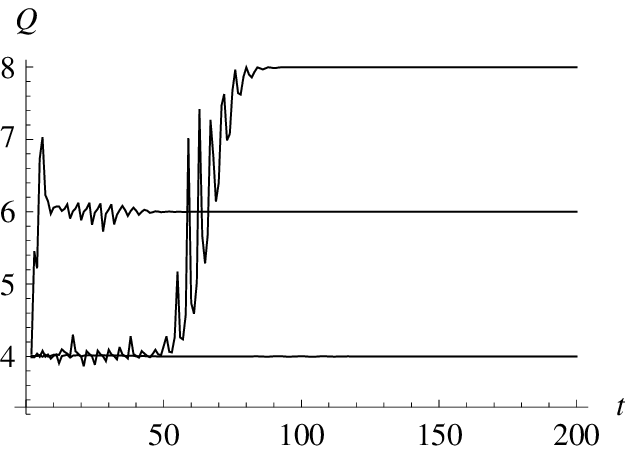}
    \hspace{1cm}
    \caption{Convergence in the $L_2$-norm indicating 4th, 6th and 8th
      order convergence for the corresponding finite difference
      operators. The convergence factor $Q$ is calculated by
      \mbox{$Q=\log_2\frac{ \| \phi^{low}-\phi^{med}\|}{\|
          \phi^{med}-\phi^{high}\|}$.} \label{fig:3}}
  \end{minipage}\hfill
  \begin{minipage}[t]{0.47\textwidth}
    \centering 
    \psfrag{t}{$\tau/m$} \psfrag{err}{$\delta \phi^{\mathrm{num}}$} 
    \includegraphics[height=0.17\textheight]{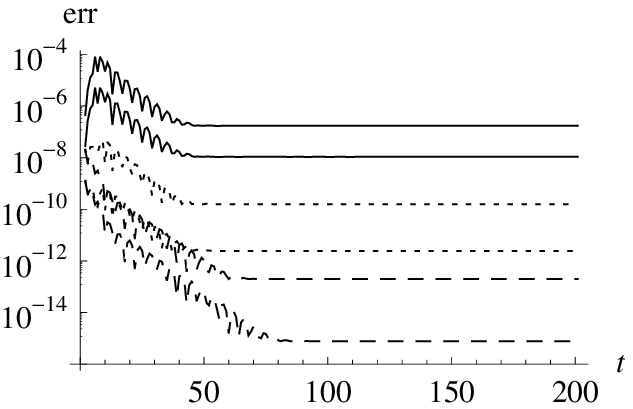}
    \caption{$L_2$-norm of the differences between numerical solutions
      obtained using different resolutions,
      i.e.~$\|\phi^{low}-\phi^{med}\|$ and $\|\phi^{med}-\phi^{high}\|$. Solid
      curves denote errors for 4th order finite difference operators,
      pointed curves for 6th order and dashed curves for 8th order.
      \label{fig:4}}
  \end{minipage}
\end{figure}

The convergence of the code can be seen in Figure \ref{fig:3}. For
this plot, a three level convergence analysis has been performed with
505, 1010 and 2020 grid cells using finite difference operators of
4th, 6th and 8th order. For the 6th and 8th order convergence tests,
we used quadruple precision because the numerical error was below
machine accuracy at late times when double precision was used. At very
late times, convergence is lost due to accumulation of numerical
errors (see the bending down of the curves in Figure \ref{fig:9} at
late times).

Figure \ref{fig:4} shows the $L_2$-norm of the relative errors between
the above mentioned simulations in different resolutions and finite
differencing orders. Using an 8th order accurate finite differencing
improves the final error by 8 orders of magnitude in the high
resolution runs confirming that high accuracies can be achieved with
high order finite differencing in combination with a relatively small
amount of grid points.

\begin{figure}[ht]
  \centering 
  \psfrag{v}{$v_\pm$} \psfrag{rho}{$\rho$} 
  \includegraphics[height=0.17\textheight]{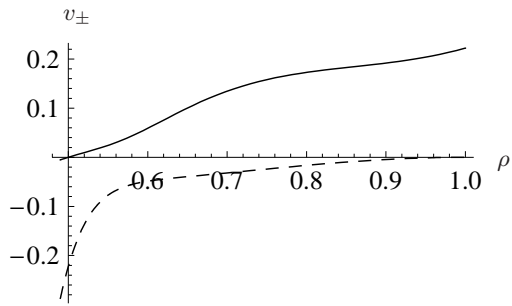}
  \caption{The dashed and the solid curves denote coordinate speeds of
    in- and outgoing characteristics respectively, for the foliation
    characterized by (\ref{eq:resc}). The horizon is located at
    $\rho=0.5$. We see that the characteristic speeds are less than
    unity which allows us to choose a large Courant factor. We observe
    that there is little variation of the speeds over the grid which
    has infinite physical extent. \label{fig:5}}
\end{figure}

The simulation domain in the compactifying coordinate reads
$\rho\in[0.495,1]$. The Courant factor $\triangle t/\triangle \rho$ in
the simulations has been chosen to be $4$. We can choose a Courant
factor that is larger than 1 because the coordinate speeds of
characteristics are less than $1$.  For outgoing and ingoing
characteristics, the characteristic speeds with respect to our
foliation are $v_\pm=\pm\balpha^2-\bbeta$. Figure \ref{fig:5} shows
the characteristic speeds along the grid for $K=1$. The allowed
Courant factor depends on the value of $K$ as $K$ is related to the
coordinate speed of characteristics. For example, the coordinate speed
of outgoing characteristics at $\scri^+$ reads in our foliation
$v_+|_{\scri^+} = 2 K^2/9$. Therefore, a large value for $K$ implies a
small Courant factor \cite{Zeng07}. We plot in Figure \ref{fig:5} the
speeds for $K=1$ and see that $v_\pm<1$ which allows $\triangle
t/\triangle \rho > 1$.  We also see that the ingoing characteristic
speed vanishes at $\scri^+$ as expected, correspondingly we do not
need to impose outer boundary conditions.  Note also that there is
relatively little variation of the characteristic speeds over this
grid which has infinite physical extent.

In the studies presented below, we choose $m=1/2, c=1$ and $K=1$. 
\subsection{Quasinormal mode ringing}
\begin{figure}[ht]
  \centering 
  \begin{minipage}[t]{0.47\textwidth}
    \psfrag{logphi}{$|\phi_\rho|$}
    \psfrag{t}{$\tau/m$}
    \includegraphics[height=0.21\textheight]{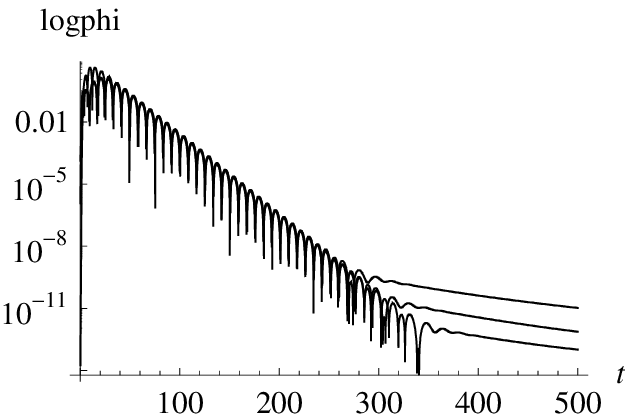}
    \hspace{1cm}
    \caption{Quasi normal mode ringing of Schwarzschild spacetime
      excited by a Gaussian gravitational perturbation with $l=2$
      measured by three observers. Looking to the tail part of the
      signal, the observers are located from top to bottom at
      $\scri^+$, $r\approx 18m$ and $r\approx 2.5m$. We see that the
      ringing phase ends at different times for different observers
      and goes into a polynomial decay. \label{fig:6}}
  \end{minipage}\hfill
  \begin{minipage}[t]{0.47\textwidth}
    \centering 
    \psfrag{ph}{$\phi$} 
    \includegraphics[height=0.17\textheight]{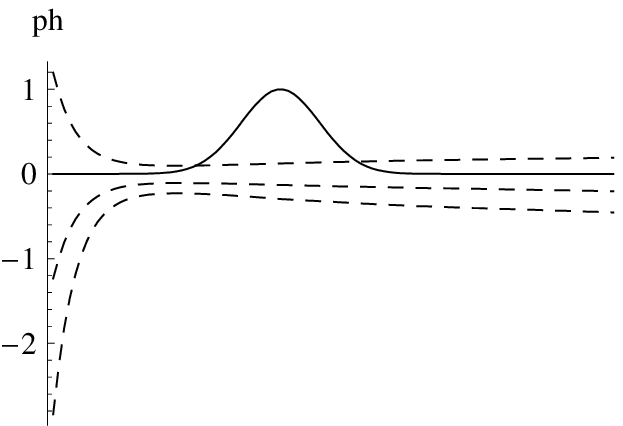}
    \caption{The numerical solution on the grid during the quasinormal
      ringing phase plotted at different times and rescaled for
      visibility. The solid curve is just the initial data. The dashed
      curves are from top to bottom: \mbox{ $\{10\times
        \phi(80m,\rho),100\times \phi(106m,\rho), 1000\times
        \phi(122m,\rho)\}$.} We see that the wave package is
      distributed quite uniformly across the grid.
    \label{fig:7}}
  \end{minipage}
\end{figure}

Figure \ref{fig:6} shows the gravitational perturbations of a
Schwarzschild black hole in a double logarithmic plot with respect to
three observers located at $r\approx \{2.5m,18m\}$ and at $\scri^+$.
We see that, as expected, the quasinormal mode frequencies are the
same for each observer. It does not matter for the calculation of
these frequencies at what distance from the black hole the
perturbations are measured. The end of the quasinormal ringing phase,
however, is observer-dependent. The polynomially decaying part of the
solution starts earlier for an observer at $\scri^+$ than for a nearby
observer. This suggests that for the calculation of quasi-normal mode
frequencies, one may get better results by using the waveform as
measured by a nearby observer.

We expect the quasinormal mode ringing to have the form
\be \label{eq:qnm} \phi_\rho(\tau) = a_\rho \,e^{-\omega_2 \tau}
\sin(\omega_1 \tau + \varphi_\rho).\ee Here, $\omega_1$ and $\omega_2$
are the mode frequencies, $a_\rho$ is the amplitude and $\phi_\rho$ is
the phase of the wave signal. In contrast to the mode frequencies, the
amplitude and the phase do depend in general on the observer's
location.

To demonstrate the efficiency of our method, we calculate the
quasinormal mode frequencies during the ringing phase using various
resolutions. The frequencies are compared to the result from Leaver's
continued fraction method \cite{Leaver:1985ax}. For our choice of
black hole mass of $m=1/2$ they read
$(\omega_1,\omega_2)=(0.747343,0.177925)$. Table \ref{table:qnm} shows
the measured frequencies for various grid resolutions for 4th and 6th
order finite differencing operators along with the relative
errors. The relative error $\delta \omega^{(2p)}_i$ for
$\omega^{\mathrm{num}(2p)}_i$ in a numerical simulation with spatial
finite differencing of order $2p$ is calculated by $\delta \omega^{(2p)}_i =
|\omega^{\mathrm{num}(2p)}_i-\omega_i|/\omega_i$. The fitting is
performed using a simple least squares method on the interval
$\tau\in[80m,180m]$ with an initial guess of the frequencies.

\begin{table}[ht]
\begin{tabular}{|c|c|c|c|c|c|c|c|c|}
\hline
Cells & $\omega^{\mathrm{num}(4)}_1$ & $\omega^{\mathrm{num}(4)}_2$ &
 $\delta \omega^{(4)}_1$ & $\delta \omega_2^{(4)}$ & $\omega^{\mathrm{num}(6)}_1$ & 
$\omega^{\mathrm{num}(6)}_2$ & $\delta \omega_1^{(6)}$ & $\delta \omega_2^{(6)}$ \\
\hline 25\footnote{For the runs with 25 grid cells, a relatively high 
dissipation parameter of $\epsilon=0.2$ has been used.} 
& 0.825455 & 0.157916 & $10^{-2}$ & $1.1\times 10^{-2}$ 
& 0.777778 & 0.168439 & $4.1\times 10^{-2}$ & $5.3\times 10^{-2}$ \\ 
\hline 50  & 0.746988 & 0.177897&$4.7\times 10^{-4}$ & $1.5\times 10^{-4}$
& 0.747341 & 0.177917 & $2.3\times 10^{-6}$ & $4.4\times 10^{-5}$ \\ 
\hline 100 & 0.747325 & 0.177927&$2.4\times 10^{-5}$ & $1.1\times 10^{-5}$
& 0.747344 & 0.177923 & $1.9\times 10^{-6}$ & $9.5\times 10^{-6}$ \\ 
\hline 200 & 0.747343 & 0.177924&$2.6\times 10^{-7}$ & $4.9\times 10^{-6}$
& 0.747344 & 0.177924 & $1.8\times 10^{-6}$ & $5.4\times 10^{-6}$ \\ 
\hline\end{tabular}
  \caption{Numerical quasinormal mode frequencies
    $\omega_i^{\mathrm{num}(2p)}$ in various resolutions measured at
    $r\approx 2.5m$ and the relative errors $\delta \omega^{(2p)}_i$
    calculated in a numerical simulation with spatial finite
    difference operators of the order $2p$. We perform a simple
    least square fit to (\ref{eq:qnm}) with initial guess. We see that
    even with low resolutions, we are able to measure the
    quasinormal-mode frequencies with a high accuracy indicating the
    efficiency of our method.
\label{table:qnm}}
\end{table}

We see that already with 50 cells we are able to calculate the
quasinormal mode frequencies with a relative error of about $10^{-4}$
using 4th order finite differences and $10^{-6}$ using 6th order
finite differences. The use of higher order accurate finite
differencing does not increase the computational cost
significantly. We observed only about a $5\%$ increase in the run time
for the 6th order finite differencing compared to the 4th order
one. We note that for a relative error at the order of $10^{-6}$, the
accuracy in the frequency is rather dominated by the fit than by
numerics. This explains, in part, why the relative error in the
simulation with 200 grid cells seems smaller in the 4th order case than
in the 6th order case. Using 8th order finite differencing gives
similar results as the 6th order one and does not increase the
accuracy in the frequencies considerably.

The fact that using a very small number of cells is already sufficient
for such accuracy in the frequencies demonstrates the efficiency of
the hyperboloidal approach.  An explanation for this efficiency can be
found in the way the solution looks like. Figure \ref{fig:7} shows the
solution at three different times during the quasinormal ringing
phase. We see that the initially localized perturbation is distributed
uniformly on the grid. This uniform distribution implies that there is
no need to use a high number of grid points to resolve the outgoing
wave package. For accuracy, it is more efficient to use a high order
scheme with a small number of grid points in this case. In evolutions
with Cauchy-type foliations, however, the outgoing wave package is a
long wave-train with the typical quasinormal ringing form that
requires good resolution. We note that the observed broadening in the
hyperboloidal approach is stronger for larger values of $K$. For
smaller values we get a similar localization like when Cauchy-type
foliations are used. This is because $K$ can be regarded as a measure
of how close the behavior of hyperboloidal surfaces is to being
characteristic \cite{Zeng07}.

\subsection{Polynomial decay}
\begin{figure}[ht]
  \centering 
  \begin{minipage}[t]{0.47\textwidth}
    \flushright 
    \psfrag{t}{$\tau/m$} \psfrag{ph}{$ |\phi_\rho|$}
    \includegraphics[height=0.2\textheight]{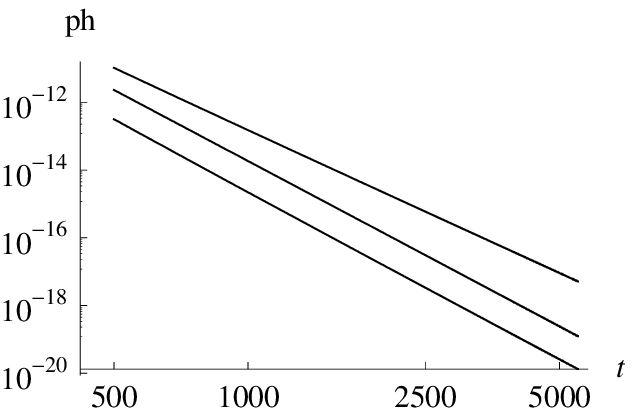}
    \hspace{1cm}
    \caption{The polynomially decaying part of the solution in a
      double-logarithmic plot starting from $\tau=500m$. The upper
      curve gives the solution at null infinity, the middle one at
      $r\approx 60 m$ and the lower one at the future horizon
      $\mathcal{H}^+$. It can be seen that the decay of the solution
      along $\scri^+$ is slower than near the black
      hole. \label{fig:8}}
  \end{minipage}\hfill
  \begin{minipage}[t]{0.47\textwidth}
    \centering 
    \psfrag{tau}{$\tau/m$} \psfrag{lpi}{$p_\rho$}
    \includegraphics[height=0.21\textheight]{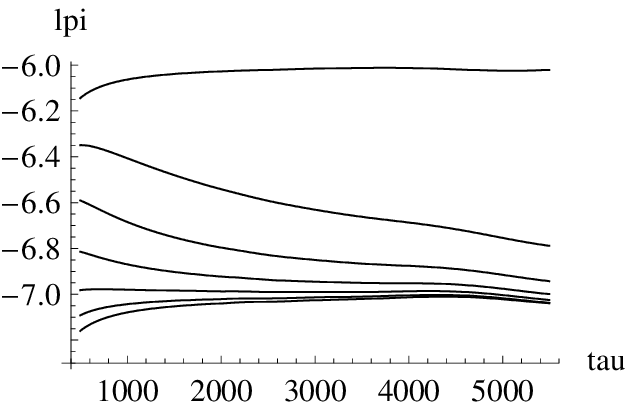}
    \caption{The local power index $p_\rho(\tau)$ for observers
      located from top to bottom at $\{\scri^+, 800m, 260m,
      110m, 50m, 25m\}$. We see that after about $\tau=4500m$ the
      local power index for nearby observers bends down. This behavior
      is seen earlier for lower resolutions and is only a numerical
      effect indicating loss of convergence due to loss of
      accuracy. \label{fig:9}}
  \end{minipage}
\end{figure}

Figure \ref{fig:8} shows the polynomially decaying part of the
solution starting from $\tau=500m$ as seen at future null infinity,
$\scri^+$, at $r\approx 60m$ and at the future horizon,
$\mathcal{H}^+$, in an evolution with 4040 grid cells. We observe that
the scattered perturbations are stronger for the observer at $\scri^+$
and also the slope of the curves are different, indicating that the
decay rate for a far away observer is less than the decay rate for a
close by observer. An intuitive explanation for this difference is
given by the picture that the tail is generated by backscattering off
the nonvanishing curvature of the background spacetime. An observer
who is farther away sees more scattering than a nearby
observer. Therefore, the amount of scattered waves measured by a
nearby observer is less than for a farther one, consequently the
scattered part of the signal is ``stronger'' for the far away observer
(modulo the natural fall-off) and decays slower. 
The natural fall-off is not seen in Figure \ref{fig:8} as the variable
we plot corresponds to the real part of $\Psi_4$ rescaled with $r$.

We see from Figures \ref{fig:6} and \ref{fig:8} that we are able to
follow the solution for more than 20 orders of magnitude and calculate
the predicted decay rate with high accuracy. To achieve such accuracy,
we use 8th order finite difference operators. The exponentially
decaying (quasinormal mode ringing) part of the solution is calculated
in quadruple precision. The reason for choosing 8th order accurate
finite differencing is due to the very small effects that we are
trying to measure. Note that for any given finite difference order,
there are ghost potentials which may be mistaken as tail solutions
\cite{Ching95}. We have tested that the tail seen in Figure
\ref{fig:8} is not a numerical artefact by comparing solutions in
various resolutions.

An accurate picture of the decay rates can be obtained from Figure
\ref{fig:9}. Here we plot, for various observers, the local power
index defined as \be\label{eq:exponent} p_\rho(\tau) = \frac{d\ln
  |\phi(\tau,\rho)|}{d\ln \tau}. \ee The function $p_\rho(\tau)$
becomes asymptotically the exponent of the polynomial decay of the
solution.  At late times, our numerically measured decay rate becomes
$-6.01$ for an observer at $\scri^+$ and $-7.002$ for an observer near
$i^+$, in accordance with the well-known prediction of $2^{-(2l+2)}$
at $\scri^+$ and of $2^{-(2l+3)}$ near $i^+$ \cite{Price72} (see also
\cite{Cunningham78, Cunningham79, Leaver:1986gd, Gundlach94a,
  Gundlach94b, Barack:1998bv, Barack:1998bw}).\pagebreak

\section{Discussion}
In this paper we have revisited the problem of scattering pulses of
gravitational waves off a Schwarzschild black hole with the aim of
getting guidance in the development of a hyperboloidal approach to
numerical relativity. We have used constant mean curvature slices on
the background of a Schwarzschild spacetime \cite{Brill80,
  MalecMurch03} in a scri-fixing gauge \cite{Zeng07} to numerically
solve a hyperboloidal initial value problem \cite{Zeng08a} for the
Bardeen-Press equation. The numerical study of gravitational
perturbations at null infinity follows, in some sense, the spirit of
\cite{Husa:2001pk,Campanelli:2000in,Zlochower:2003yh} who studied
compactified null foliations.

Motivation for taking a hyperboloidal approach comes from the
well-known shortcomings of existing methods based on Cauchy-type and
characteristic foliations for this problem. The drawbacks of using
Cauchy-type foliations can be summarized as follows: (i) waste of
computational resources when one is interested in the waveform as seen
by far away observers, (ii) spurious reflections from the outer
boundary contaminating the waveform in long time evolutions due to
unknown boundary data, (iii) not being able to read off the waveform
at null infinity. Using a characteristic foliation solves these
problems, but leads to new difficulties: (i) the causal relationship
between the surfaces seems to require a two-stage approach
\cite{Campanelli:2000in,Husa:2001pk}, (ii) choice of variables for the
numerical simulations is not straightforward due to loss of
resolution, and most importantly, (iii) extension beyond spherical
symmetry is very complicated. Especially in highly dynamical
situations where formation of caustics is expected \cite{Friedrich83},
it is not yet clear how the characteristic approach can be implemented
\cite{Winicour05}.

The hyperboloidal approach hopes to combine the best properties of
currently used approaches in the field, namely the flexibility of
Cauchy-type surfaces and the asymptotic behavior of characteristic
surfaces, and avoids their problems -- however it is also much less
developed so far.

We believe that the problem of treating gravitational perturbations on
a fixed spherically symmetric background, while rather simple, is an
important step in the development of algorithms for the hyperboloidal
initial value problem. We have chosen to solve the Bardeen-Press
equation describing perturbations of Schwarzschild spacetime in terms
of the Weyl component $\psi_4$, partly because this naturally
generalizes to the much more interesting axisymmetric case via the
Teukolsky equation.

We demonstrated in this paper that compactification along
hyperboloidal surfaces using the Teukolsky formalism on the background
of a Schwarzschild spacetime leads to regular equations. This regular
behavior can be expected on the background of any weakly
asymptotically simple spacetime (in the sense of Penrose
\cite{Penrose63, Penrose65}). The resulting equations can be solved
numerically to study quasinormal mode ringing very efficiently. The
tail part can be calculated accurately with sufficient numerical
resolution. There are no reflections from the numerical outer
boundary. The only approximation we use to calculate the gravitational
waveform beside the linearization is the numerical discretization of
the equations.  The observed uniform distribution of outgoing wave
packages along the grid and the numerical experiments we performed
suggest that the use of high order methods to solve the linear
perturbation equations lead to an efficient code that requires a
moderate number of grid points while achieving good accuracy in the
results.

We want to conclude with a brief discussion of open problems for
future work.  It may be interesting to compare the approach for
curvature-based and metric perturbations using the Chandrasekhar
transformation along the lines of \cite{Campanelli:1998uh}. Current
work is under way to study metric perturbations using the
Regge-Wheeler-Zerilli formalism in a hyperboloidal setting
\cite{Moncrief}. The hyperboloidal approach to study gravitational
perturbations at null infinity should generalize rather
straightforwardly to Kerr spacetime, for example by using the
foliations presented in \cite{Zeng07}.

The difficulty of generalization to the fully nonlinear Einstein
equations remains unclear. In contrast to the test field case, the
conformally rescaled Einstein equations are not explicitly
regular. There, a scri-fixing gauge can be constructed in a
hyperboloidal initial value problem for the generalized harmonic
reduction of the Einstein equations such that the formally singular
terms attain regular limits at null infinity
\cite{Zenginoglu:2008pw}. The open question is whether such limits can
be calculated numerically in a stable way. The asymptotic behavior of
the gauge functions (\ref{eq:scrifix}) need to be taken into account
if a regularization along the lines of \cite{Zenginoglu:2008pw} needs
to be done for the ADM or BSSN systems based on scri-fixing. A very
promising alternative is to adopt an elliptic-hyperbolic formulation
of the problem, and apply boundary conditions for gauge quantities as
an elliptic Dirichlet problem, as suggested in
\cite{Andersson:2002gn}.

\acknowledgments 

This work was supported in part by the NSF grant PHY0801213 to the
University of Maryland and by the DFG grant SFB/Transregio~7
``Gravitational Wave Astronomy''. S. Husa is a VESF fellow of the
European Gravitational Observatory (EGO).  D N\'u\~{n}ez is grateful
to Luciano Rezzolla for warm hospitality during his stay at the AEI,
and acknowledges DAAD, DGAPA-UNAM and CONACyT grants for partial
support.

\appendix*
\section{Symmetric hyperbolic form of the Bardeen-Press equation}
We solve the Bardeen-Press equation in first order symmetric
hyperbolic form using a spherically symmetric CMC-foliation of
Schwarzschild spacetime with the gauge functions given in
(\ref{eq:gaugephy}). The relation $\alpha=1/\gamma$ simplifies the
calculation, so we will assume it in the following. In such a gauge,
the spinor coefficients take the form
\[ \rho_s = \frac{\alpha^2-\beta}{2\,\alpha^2 r}, \qquad \mu_s =
\frac{\alpha^2+\beta}{r}.\] 

To bring the radial part of the Bardeen-Press equation
(\ref{eq:pertPhi1KSf}) to first order form, we introduce the auxiliary
variables \be\label{eq:aux} \psi:= \partial_r \phi,
\quad\mathrm{and}\quad \pi :=
\frac{1}{\alpha^2}(\partial_t\phi-\beta\,\partial_r \phi). \ee Then
the following linear, symmetric hyperbolic system of evolution
equations is obtained from (\ref{eq:optrPhi1}), (\ref{coef_fo1})
\begin{eqnarray}\label{eq:firstorder}
  \partial_t \phi &=&\alpha^2 \,\pi + \beta\psi, \nonumber
  \\ \partial_t \psi &=&\partial_r\left(\alpha^2 \pi +
  \beta\psi\right), \\ \partial_t \pi &=&
  \partial_r\left(\alpha^2\psi+\beta\pi\right)+A_\pi\,\pi+A_\psi\,\psi+\left(A_\phi-\frac{\lambda}{r^2}\right)\phi,
  \nonumber
\end{eqnarray}
with $\lambda = \left(l-1\right)\left(l+2\right)$ and
\begin{eqnarray}
  A_\pi &=&-4\,(\alpha^2+\beta)\,\partial_r\left(
  \ln\frac{\alpha}{r}\right), \nonumber\\ \label{eq:ABC} 
 A_\psi &=&A_\pi -
  4\,\alpha^2\,\partial_r\left(\frac{\beta}{\alpha^2}\right), \\ 
 A_\phi &=&-\frac1r\partial_r\left(\alpha^2 -\frac{\beta^2}{\alpha^2}\right)
  + \frac4r\left(\alpha^2\partial_r\left(\frac{\beta}{\alpha^2}\right)
  + \left(\alpha^2 -\frac{\beta^2}{\alpha^2}\right)\partial_r\left(
  \ln\frac{\alpha}{r}\right)\right) +
  \frac{2\left(\alpha^2+\beta\right)^2}{r^2}
  \partial_r\left(\frac{\partial_r\left(r^2\left(1-\frac{\beta}{\alpha^2}\right)\right)}{\alpha^2+\beta}\right). \nonumber
\end{eqnarray}

\bibliographystyle{unsrt}
\bibliography{references}

\begin{thebibliography}{10}

\bibitem{Pretorius:2005gq}
Frans Pretorius.
\newblock {Evolution of binary black hole spacetimes}.
\newblock {\em Phys. Rev. Lett.}, 95:121101, 2005.

\bibitem{Campanelli:2005dd}
Manuela Campanelli, C.~O. Lousto, P.~Marronetti, and Y.~Zlochower.
\newblock {Accurate evolutions of orbiting black-hole binaries without
  excision}.
\newblock {\em Phys. Rev. Lett.}, 96:111101, 2006.

\bibitem{Baker:2005vv}
John~G. Baker, Joan Centrella, Dae-Il Choi, Michael Koppitz, and James van
  Meter.
\newblock {Gravitational wave extraction from an inspiraling configuration of
  merging black holes}.
\newblock {\em Phys. Rev. Lett.}, 96:111102, 2006.

\bibitem{Hannam:2007wf}
Mark Hannam, Sascha Husa, Bernd Br{\"u}gmann, and Achamveedu Gopakumar.
\newblock Comparison between numerical-relativity and post-{N}ewtonian
  waveforms from spinning binaries: the orbital hang-up case.
\newblock 2007.

\bibitem{Hinder:2008kv}
Ian Hinder, Frank Herrmann, Pablo Laguna, and Deirdre Shoemaker.
\newblock {Comparisons of eccentric binary black hole simulations with
  post-Newtonian models}.
\newblock 2008.

\bibitem{Campanelli:2008nk}
Manuela Campanelli, Carlos~O. Lousto, Hiroyuki Nakano, and Yosef Zlochower.
\newblock {Comparison of Numerical and Post-Newtonian Waveforms for Generic
  Precessing Black-Hole Binaries}.
\newblock 2008.

\bibitem{Boyle:2008ge}
Michael Boyle et~al.
\newblock {High-accuracy numerical simulation of black-hole binaries:
  Computation of the gravitational-wave energy flux and comparisons with
  post-Newtonian approximants}.
\newblock 2008.

\bibitem{Baker:2008mj}
John~G. Baker et~al.
\newblock {Mergers of non-spinning black-hole binaries: Gravitational radiation
  characteristics}.
\newblock {\em Phys. Rev.}, D78:044046, 2008.

\bibitem{Pollney:2007ss}
Denis Pollney et~al.
\newblock Recoil velocities from equal-mass binary black-hole mergers: a
  systematic investigation of spin-orbit aligned configurations.
\newblock {\em Phys. Rev. D}, 76:124002, 2007.

\bibitem{Hannam:2007ik}
Mark Hannam, Sascha Husa, Ulrich Sperhake, Bernd Br{\"u}gmann, and Jos{\'e}~A.
  Gonz{\'a}lez.
\newblock Where post-{N}ewtonian and numerical-relativity waveforms meet.
\newblock 2007.

\bibitem{Boyle:2007ft}
Michael Boyle et~al.
\newblock {High-accuracy comparison of numerical relativity simulations with
  post-Newtonian expansions}.
\newblock {\em Phys. Rev.}, D76:124038, 2007.

\bibitem{Pazos2006}
Enrique Pazos, Ernst~Nils Dorband, Alessandro Nagar, Carlos Palenzuela, Erik
  Schnetter, and Manuel Tiglio.
\newblock How far away is far enough for extracting numerical waveforms, and
  how much do they depend on the extraction method?
\newblock {\em (submitted to Class. Quantum Grav.)}, 2006.

\bibitem{Kocsis:2007zz}
Bence Kocsis and Abraham Loeb.
\newblock {Distortion of gravitational-wave packets due to their self-gravity}.
\newblock {\em Phys. Rev.}, D76:084022, 2007.

\bibitem{Campanelli:1998uh}
Manuela Campanelli, William Krivan, and Carlos~O. Lousto.
\newblock {The imposition of Cauchy data to the Teukolsky equation II:
  Numerical comparison with the Zerilli-Moncrief approach to black hole
  perturbations}.
\newblock {\em Phys. Rev.}, D58:024016, 1998.

\bibitem{Baker:2001sf}
John~G. Baker, Manuela Campanelli, and Carlos~O. Lousto.
\newblock {The Lazarus project: A pragmatic approach to binary black hole
  evolutions}.
\newblock {\em Phys. Rev.}, D65:044001, 2002.

\bibitem{PazosAvalos:2004rp}
Enrique Pazos-Avalos and Carlos~O. Lousto.
\newblock {Numerical integration of the Teukolsky equation in the time domain}.
\newblock {\em Phys. Rev.}, D72:084022, 2005.

\bibitem{Husa:2001pk}
Sascha Husa, Yosef Zlochower, Roberto Gomez, and Jeffrey Winicour.
\newblock {Retarded radiation from colliding black holes in the close limit}.
\newblock {\em Phys. Rev.}, D65:084034, 2002.

\bibitem{Campanelli:2000in}
Manuela Campanelli, Roberto Gomez, Sascha Husa, Jeffrey Winicour, and Yosef
  Zlochower.
\newblock {The close limit from a null point of view: The advanced solution}.
\newblock {\em Phys. Rev.}, D63:124013, 2001.

\bibitem{Zlochower:2003yh}
Yosef Zlochower, Roberto Gomez, Sascha Husa, Luis Lehner, and Jeffrey Winicour.
\newblock {Mode coupling in the nonlinear response of black holes}.
\newblock {\em Phys. Rev.}, D68:084014, 2003.

\bibitem{Bizon:2007qz}
Piotr Bizon, Tadeusz Chmaj, and Andrzej Rostworowski.
\newblock {On asymptotic stability of the Skyrmion}.
\newblock {\em Phys. Rev.}, D75:121702, 2007.

\bibitem{Bizon:2007xa}
Piotr Bizon, Tadeusz Chmaj, and Andrzej Rostworowski.
\newblock {Late-time tails of a Yang-Mills field on Minkowski and Schwarzschild
  backgrounds}.
\newblock {\em Class. Quant. Grav.}, 24:F55, 2007.

\bibitem{Baker:1999sj}
John~G. Baker et~al.
\newblock {Nonlinear and Perturbative Evolution of Distorted Black Holes. II.
  Odd-parity Modes}.
\newblock {\em Phys. Rev.}, D62:127701, 2000.

\bibitem{Baker:2002qf}
John~G. Baker, Manuela Campanelli, C.~O. Lousto, and R.~Takahashi.
\newblock {Modeling gravitational radiation from coalescing binary black
  holes}.
\newblock {\em Phys. Rev.}, D65:124012, 2002.

\bibitem{Baker:2003ds}
John~G. Baker, Manuela Campanelli, Carlos~O. Lousto, and R.~Takahashi.
\newblock {Coalescence remnant of spinning binary black holes}.
\newblock {\em Phys. Rev.}, D69:027505, 2004.

\bibitem{Campanelli:2005ia}
Manuela Campanelli, Bernard~J. Kelly, and Carlos~O. Lousto.
\newblock The {L}azarus project {II}: Space-like extraction with the
  quasi-{K}innersley tetrad.
\newblock {\em Phys. Rev. D}, 73:064005, 2006.

\bibitem{Bardeen:1973xb}
J.~M. Bardeen and W.~H. Press.
\newblock {Radiation fields in the {Schwarzschild} background}.
\newblock {\em J. Math. Phys.}, 14:7--19, 1973.

\bibitem{Allen:2004js}
Elspeth~W. Allen, Elizabeth Buckmiller, Lior~M. Burko, and Richard~H. Price.
\newblock {Radiation tails and boundary conditions for black hole evolutions}.
\newblock {\em Phys. Rev.}, D70:044038, 2004.

\bibitem{Dafermos:2004wt}
Mihalis Dafermos and Igor Rodnianski.
\newblock {A note on boundary value problems for black hole evolutions}.
\newblock 2004.

\bibitem{Leaver:1986gd}
Edward~W. Leaver.
\newblock {Spectral decomposition of the perturbation response of the
  Schwarzschild geometry}.
\newblock {\em Phys. Rev.}, D34:384--408, 1986.

\bibitem{Purrer:2004nq}
Michael P{\"u}rrer, Sascha Husa, and Peter~C. Aichelburg.
\newblock {News from critical collapse: Bondi mass, tails and quasinormal
  modes}.
\newblock {\em Phys. Rev.}, D71:104005, 2005.

\bibitem{Zeng08a}
An{\i}l Zengino\u{g}lu.
\newblock A hyperboloidal study of tail decay rates for scalar and {Yang-Mills}
  fields.
\newblock {\em Class. Quant. Grav.}, 25:175013, 2008.

\bibitem{Teu73}
S.~A. Teukolsky.
\newblock Perturbations of a rotating black hole {I}. {F}undamental equations
  for gravitational, electromagnetic, and neutrino-field perturbations.
\newblock {\em Astroph. J.}, 185:635--647, 1973.

\bibitem{Bishop05}
Liebrecht~R. Venter and Nigel~T. Bishop.
\newblock Numerical validation of the kerr metric in {Bondi-Sachs} form.
\newblock {\em Phys. Rev.}, D73:084023, 2006.

\bibitem{Pretorius98}
Frans Pretorius and Werner Israel.
\newblock Quasi-spherical light cones of the {Kerr} geometry.
\newblock {\em Class. Quant. Grav.}, 15:2289--2301, 1998.

\bibitem{Friedrich83a}
Helmut Friedrich.
\newblock Cauchy problems for the conformal vacuum field equations in general
  relativity.
\newblock {\em Comm. Math. Phys.}, 91:445--472, 1983.

\bibitem{Husa:2005ns}
Sascha Husa, Carsten Schneemann, Tilman Vogel, and Anil Zenginoglu.
\newblock {Hyperboloidal data and evolution}.
\newblock {\em AIP Conf. Proc.}, 841:306--313, 2006.

\bibitem{Andersson:2002gn}
Lars Andersson.
\newblock Construction of hyperboloidal initial data.
\newblock {\em Lect. Notes Phys.}, 604:183--194, 2002.

\bibitem{NP62}
E.~Newman and R.~Penrose.
\newblock An approach to gravitational radiation by a method of spin
  coefficients.
\newblock {\em J. Math Phys.}, 3:566--579, 1962.

\bibitem{Chandra83}
S.~Chandrasekhar.
\newblock {\em The mathematical theory of black holes}.
\newblock Clarendon Press Oxford, Oxford, UK, 1983.

\bibitem{Frolov77}
V.~Frolov.
\newblock Newman-penrose method in the general relativistic theory (in
  {Russian}).
\newblock {\em Science Academy USSR}, 96:72--180, 1977.

\bibitem{Ernst78}
F.~J. Ernst.
\newblock Coping with different languages in the null tetrad formulation of
  general relativity.
\newblock {\em J. Math Phys.}, 19:489--493, 1978.

\bibitem{Newm66}
E.~Newman and R.~Penrose.
\newblock Note on the {Bondi-Metzner-Sachs} group.
\newblock {\em J. Math Phys.}, 7:863--870, 1966.

\bibitem{Gold67}
J.~N. Goldberg, A.~J. Macfarlane, E.~Newman, F.~Rohrlich, and E.~C.~G.
  Sudarshan.
\newblock Spin s spherical harmonics and eth.
\newblock {\em J. Math Phys.}, 8:2155--2161, 1967.

\bibitem{Sarbach:2001qq}
Olivier Sarbach and Manuel Tiglio.
\newblock {Gauge invariant perturbations of Schwarzschild black holes in
  horizon-penetrating coordinates}.
\newblock {\em Phys. Rev.}, D64:084016, 2001.

\bibitem{Brill80}
Dieter~R. Brill, John~M. Cavallo, and James~A. Isenberg.
\newblock K-surfaces in the {Schwarzschild} space-time and the construction of
  lattice cosmologies.
\newblock {\em Journal of Mathematical Physics}, 21(12):2789--2796, 1980.

\bibitem{MalecMurch03}
Edward Malec and Niall~O Murchadha.
\newblock Constant mean curvature slices in the extended {S}chwarzschild
  solution and collapse of the lapse: Part {I}.
\newblock {\em Phys. Rev. D}, 68:124019, 2003.

\bibitem{Zeng07}
An{\i}l Zengino\u{g}lu.
\newblock {Hyperboloidal foliations and scri-fixing}.
\newblock {\em Class. Quant. Grav.}, 25:145002, 2008.

\bibitem{KreissOliger73}
H.~O. Kreiss and J.. Oliger.
\newblock {\em Methods for the approximate solution of time dependent
  problems}.
\newblock International Council of Scientific Unions, World Meteorological
  Organization, Geneva, 1973.

\bibitem{Leaver:1985ax}
E.~W. Leaver.
\newblock {An analytic representation for the quasi normal modes of {Kerr}
  black holes}.
\newblock {\em Proc. Roy. Soc. Lond.}, A402:285--298, 1985.

\bibitem{Ching95}
E.~S.~C. Ching, P.~T. Leung, W.~M. Suen, and K.~Young.
\newblock Wave propagation in gravitational systems: Late time behavior.
\newblock {\em Phys. Rev. D}, 52(4):2118--2132, 1995.

\bibitem{Price72}
R.~Price.
\newblock Nonspherical perturbations of relativistic gravitational collapse.
  {I}. {Scalar} and gravitational perturtbations.
\newblock {\em Phys. Rev. D}, 5:2419, 1972.

\bibitem{Gundlach94a}
C.~Gundlach, R.~Price, and J.~Pullin.
\newblock Late-time behaviour of stellar collapse and explosions: {I}.
  linearized perturbations.
\newblock {\em Phys. Rev. D}, 49, 1994.

\bibitem{Gundlach94b}
C.~Gundlach, R.~Price, and J.~Pullin.
\newblock Late-time behaviour of stellar collapse and explosions: {II}.
  nonlinear evolution.
\newblock {\em Phys. Rev. D}, 49, 1994.

\bibitem{Barack:1998bv}
Leor Barack.
\newblock Late time dynamics of scalar perturbations outside black holes. {I}:
  A shell toy-model.
\newblock {\em Phys. Rev. D}, 59:044016, 1999.

\bibitem{Barack:1998bw}
Leor Barack.
\newblock Late time dynamics of scalar perturbations outside black holes. {II}:
  {S}chwarzschild geometry.
\newblock {\em Phys. Rev. D}, 59:044017, 1999.

\bibitem{Cunningham78}
C.~T. Cunningham, R.~H. Price, and V.~Moncrief.
\newblock Radiation from collapsing relativistic stars. {I}. {L}inearized
  odd-parity radiation.
\newblock {\em Astrophys. J.}, 224:643, 1978.

\bibitem{Cunningham79}
C.T. Cunningham, R.H. Price, and V.~Moncrief.
\newblock Radiation from collapsing relativistic stars. {II}. linearized
  even-parity radiation.
\newblock {\em Astrophys. J.}, 230:870--892, 1979.

\bibitem{Friedrich83}
Helmut Friedrich and John Stewart.
\newblock Characteristic initial data and wavefront singularities in general
  relativity.
\newblock {\em Proc. R. Soc.}, A385:345--371, 1983.

\bibitem{Winicour05}
Jeffrey Winicour.
\newblock Characteristic evolution and matching.
\newblock {\em Living Rev. Relativity}, 8:10, 2005.
\newblock [Online article].

\bibitem{Penrose63}
Roger Penrose.
\newblock Asymptotic properties of fields and space-times.
\newblock {\em Phys. Rev. Lett.}, 10:66--68, 1963.

\bibitem{Penrose65}
Roger Penrose.
\newblock Zero rest-mass fields including gravitation: Asymptotic behaviour.
\newblock {\em Proc. Roy. Soc. Lond.}, A284:159--203, 1965.

\bibitem{Moncrief}
Vincent Moncrief.
\newblock {P}rivate communication.

\bibitem{Zenginoglu:2008pw}
An{\i}l Zengino\u{g}lu.
\newblock {Hyperboloidal evolution with the {Einstein} equations}.
\newblock {\em Class. Quant. Grav.}, 25:195025, 2008.

\end{thebibliography}

\end{document}